# A Review of Technical Problems when Conducting an Investigation in Cloud Based Environments

Manuel Jesús Rivas Sández

# Abstract

Cloud computing is a relatively new technology which is quickly becoming one of the most important technological advances for computer science. This technology has had a significant growth in recent years. It is now more affordable and cloud platforms are becoming more stable. Businesses are successfully migrating their systems to a cloud infrastructure, obtaining technological and economic benefits. However, others still remain reluctant to do it due to both security concerns and the loss of control over their infrastructures and data that the migration entails.

At the same time that new technologies progress, its benefits appeal to criminals too. They can not only steal data from clouds, but they can also hide data in clouds, which has provoked an increased in the number of cybercrimes and their economic impacts. Their victims range from children and adults to companies and even countries.

On the other hand, digital forensics have negatively suffered the impact of the boom of cloud computing due to its dynamic nature. The tools and procedures that were successfully proved and used in digital investigations are now becoming irrelevant, making it an urging necessity to develop new forensics capabilities for conducting an investigation in this new environment. As a consequence of these needs a new area has emerged, Cloud Forensics, which is the result of the intersection between cloud computing and digital forensics.

## 1. Introduction

Cloud computing is a relatively new technology which has experienced a rapid growth in recent years while offering elasticity, flexibility and services on demand. According to the Fourth Annual Future of Cloud Computing Survey results, cloud computing continued rising in 2014 with 45 percent of those surveyed declaring they run their companies from the cloud. The sharp increase in its use, however, is not exempt from critical issues, recent studies have identified a number of them related to security (Armbrust, et al., 2009) (Ming and Yongsheng, 2012) (Sinha and Khreisat, 2014). A study also revealed that the main concern is that data is stored in an unknown place for the customer, being usually at any location in different countries (Sabahi, et al., 2011).

Unfortunately, cloud computing not only introduces technological and economic opportunities but also presents a better and more sophisticated environment for criminals (Yan, et al., 2011). Several researchers have claimed that cloud architecture also poses certain challenges for conducting forensics digital investigations (Taylor, et al., 2011) (Trenwith and Venter, 2013) (Thethi and Keane, 2014). Both factors, better environment for criminals and difficulties for investigators, have resulted in an augmentation in cybercrime, as it can be seen in the recently released 2014 Ponemon Institute annual report, which shows that the cost of cybercrime has increased more than 9% over the course of the year (see Figure 1). The average time to resolve an attack has also increased to 45 days, a rise of 40%.

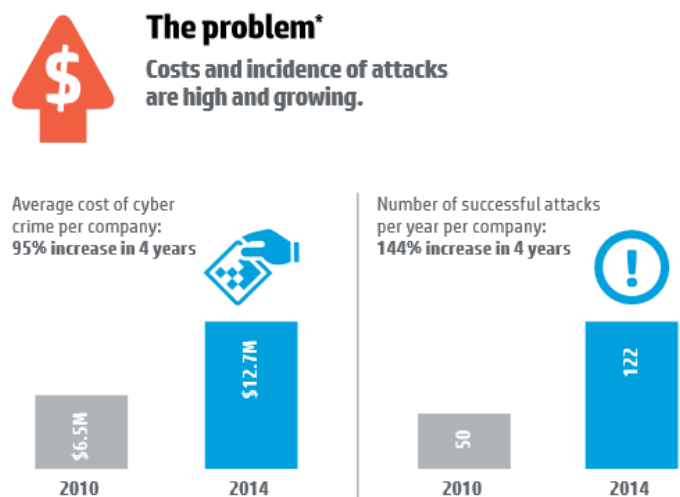

**Figure 1. The cost of cybercrime and the number of attack (2014 Ponemon Institute annual report)**

This report makes it clear that as Damshenas, et al., (2012) had previously pointed out, more attention has to be paid to cloud security and hence to forensics investigations in cloud.

In this paper, we examine the issues that can be an obstacle when conducting a cloud based investigation. Much of the work will be focused on reviewing the existing literature with the intention to provide a comprehensive analysis. The aim is to clarify the stages that need to be followed in an investigation which is conducted in the context of cloud computing, answering the follow question: Which approaches can reduce the technical complexities associated with performing forensics in cloud based environments?

## 2. An overview of Cloud Forensics

Digital Forensics is defined by the National Institute of Standards and Technology (NIST) as the application of computer investigations and analysis techniques to determine potential evidence with the aim of presenting them in court. Carstensen, et al., (2012) said that the phases of a digital investigation process consist of examining the possible evidence, preserving the findings and maintaining a strict chain of custody for the data obtained, in order to obtain reliable final evidence. It is possible to initiate a forensics investigation as a request from either a private company or law court and can be done for different reasons, such as a criminal investigation, fraudulent activities, pornography and so on. In order to perform a digital investigation in a traditional computing environment, there are different digital forensics tools that have been successfully proved, according to a 2009 Garner study; with regard to the collection and preservation data phases, Encase and FDK are the tools which are most widely accepted among the forensics community (Heiser, 2009).

Regarding the evolution of this science, Garfinkel (2010) claims that although Digital Forensics has emerged as an important and successful tool in the fight against crime, this "Golden Age of Digital Forensics" was due to finish; due to the rapid advance of the new technologies, current forensics tools and process will quickly become obsolete. The authors also state the necessity of focussing the academic research agenda on improving and updating the current tools and research process as the way to find a solution.

As it was predicted, the boom of cloud computing has introduced new challenges for digital forensics investigators due to the way in which this technology delivers its services. Studies have identified important jurisdictional issues concerning lack of international coordination, difficulties in accessing evidence, and problems to preserve the integrity of the evidence (Damshenas et al., 2012) (Aydin & Jacob, 2013). Timing is another important aspect, Thethi and Keane (2014) suggest that this is due to the almost infinite capacity of storage that Cloud Service Providers (CSPs) are offering. Apart from those challenges, the nature of cloud computing, which is based on remote storage and virtualization technologies, makes it impossible to use traditional forensics tools and methodologies (O'Shaughnessy, et al., 2013).

It was Ruan, et al., (2011) who first introduced the term of cloud forensics referring to the growing need for conducting digital investigations in cloud computing environments; they define cloud forensics as a new discipline resulting from crossing cloud computing and digital forensics, as shown in Figure 2. In the same study the researchers mention that this need was expected to rise with the growth of cloud computing.

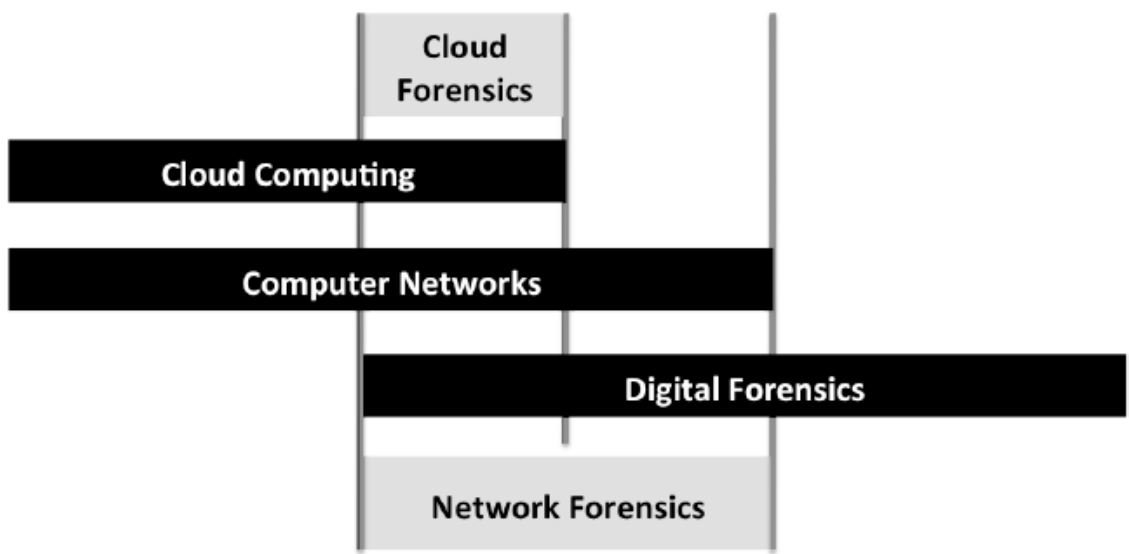

**Figure 2. Where is cloud forensics? (Ruan, et al., 2011)**

Damshenas, et al., (2012) claim that there is an urgent need to update the forensics methods, tools and techniques, and indicate that the main problem is the lack of a global cloud computing standard, which provokes confusion among the forensics investigators. Many researchers agree with that necessity, which is the reason why research to date has focused on formulating different approaches to a cloud forensics standard (Dykstra and Sherman, 2012) (Ruan and Carthy, 2013), however, the lack of procedures and forensics tools still remain a problem (Shah and Malik, 2013). NIST has also recently pointed out that there is a pressing need to develop forensics protocols, which must address the necessities of the investigators and the court system, trying not to alter the way in which the CSPs are offering their service or at least minimize this disruption (NIST, 2014).

## 3. Dimensions of Cloud Forensics

At this point, saying that computer forensics investigations are compulsory in the cloud in the interest of both assessing risk properly and to establish standards may seem pretty obvious.

In order to develop new techniques and tools for conducting an investigation the first essential step is to have a comprehensive view of cybercrime investigations domain; this understanding would help to define requirements and establish a standard (Ciardhuáin, 2004). This process, which was formulated trying to help in the development of forensics tools and techniques in a traditional environment, could now be extrapolated and applied to do the same in a cloud computing based environment. Hence, in the first step we need to gain a good understanding of the domain of forensics investigations in this new environment and analyse the different scenarios in which a cloud forensic investigation can be conducted.

With the intention to clarify the domain of cloud forensics investigations and allow future researchers to develop a standard protocol, Ruan, et al. (2011) propose a multidimensional model, which divides into three different areas the domain of cloud forensics: organizational, legal and technical, as shown in Figure3.

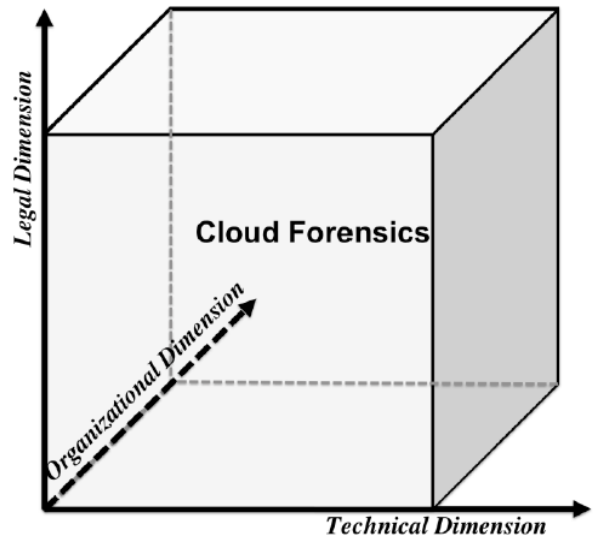

Figure 3. Cloud Forensics three-dimensional model (Ruan, et al., 2011)

The authors also emphasize how important it is to consider the cloud forensics issues as multidimensional instead of only technical issues, and define each dimension as follows: Firstly, the procedures and tools needed to conduct a forensics investigation in a cloud-based environment compound the technical dimension. These procedures need to be done without compromising the information of other tenants of the cloud service provider (CSP). Secondly, conducting an investigation in cloud implies dealing with third parties, such as CSPs; these third parties involved in the investigation process make up the organizational dimension. Finally, the legal dimension embraces the challenges in relation to different country legislations when conducting an investigation, and the preservation of the confidentiality of other tenants of the CSP.

The researchers not only propose the model but also discuss eight different issues or challenges associated with the establishment of a cloud forensics model covering the technical, organizational and legal dimensions. The challenges named by them are: forensics data acquisition, live forensics, evidence segregation, internal staffing, external dependency chains, virtualized environment, service legal agreement, multiple jurisdiction and tenancy. The first three affect the technical dimension, the next three affect to the organizational dimension and the last two affect the legal dimension. Due to constraints we are focusing only on the technical dimension in this paper.

## 4. Cloud Models and Associated Technical Dimension Challenges

It was above highlighted the necessity of analysing the wide variety of scenarios in which an investigation may be conducted, for which we need to distinguish the different models of cloud computing.

The 2011 NIST cloud computing definition classifies into three the service models available: Software as a service (SaaS), Platform as a service (PaaS) and Infrastructure as a service (IaaS). In the same definition, it mentions private clouds, public clouds and hybrid clouds among the deployment models available (Liu, et al., 2011).

Each one, either service or deployment models, have different characteristics and represent a different scenario, which makes it more difficult to define a cloud forensics standard. Zawoad and Hasan, (2013) point out one of the differences when they mention that when conducting an investigation in cloud computing, physical access to the evidence is complicated, being currently only possible in a private deployment model. Hence, due to the fact that each one presents unique features, the investigation process will vary depending on the services and deployment model of cloud computing in which the investigation is being conducted. The different cloud model also needs to be taken into consideration when analysing the different challenges associated.

In the following sections, the forensics challenges associated to the technical dimension are analysed taking into consideration the service and deployment model implemented. The analysis is also based on the three-dimensional model proposed in Ruan et al. 2011.

### 4.1 Forensics Data Collection

The first one, is the data collection process which has been named as one of the most difficult to solved (Zawoad and Hasan, 2013). This is the reason why increasing emphasis has been placed on investigating the issues related to data acquisition when conducting an investigation in a cloud based environment (Ruan, et al., 2011) (Taylor, et al., 2011) (Dykstra and Sherman, 2011). Taylor, et al., (2011) say that the impossibility of gaining access to the evidence in Hybrid and Public clouds is the main problem of the collecting data process. Ruan et al., (2011) adds that it is also important the cloud service model, as the level of control that the customer has of its own data depends on the cloud service model implemented, IaaS, SaaS or PaaS. Hence, the forensics data collection process cannot be the same in different service models, whereas in IaaS we can gain access to the virtual machine instance from the customer interface and implement some tools, in the first two we exclusively depend on the CSPs (Zawoadand Hasan, 2013).

Dykstra and Sherman (2012) conducted a practical experiment in which they used traditional digital forensics tools, Encase and FDK, successfully for acquiring forensics data in IaaS. However, they argue that too much trust in the cloud provider is needed as it was the provider instead of the investigator who had the control over the tools. The intervention of this third party may compromise the validity of the evidence in court. They also indicate that the customer management plane is the best option for forensics data acquisition in an IaaS deployment model, as the investigator can collect forensics data without asking for the assistance of the cloud provider. Hence, dependency on the CSP is not necessary. Recently, in further research Dykstra, and Sherman, (2013) presented a collection of three forensics tools. The tool which is known as Forensics Open-Stack Tools (FROST) was installed and successfully proved in an OpenStack instance (IaaS). FROST, being integrated into the management plane of the CSPs enables forensics investigators to acquire trustworthy forensics data of virtual disks, API logs and guest firewall logs. This process is developed without any dependency on the CSP. Figure 4 shows how two of the OpenStack components, Compute (Nova) and Dashboard (Horizon), have been modified to add FROST.

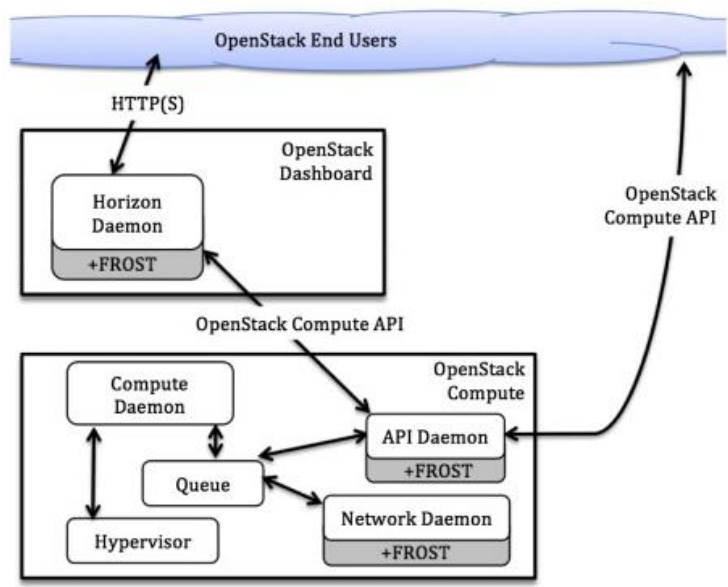

**Figure 4. Snippet of the OpenStack architecture using FROST (Dykstra and Sherman, 2013)**

Regarding the SaaS and PaaS service models, it was above highlighted the less control that users have in these services models and the limited access available in both of them. Dykstra and Sherman (2013) point out makes the process of collecting forensics data more complicated. In further research Zawoad and Hasan (2013) refer to recent studies to propose a Trusted Third Party Model in which forensics data would be collected by the cloud service provider following a set of rules previously established between the third parties implied in the process. They also argue that although this model is still immature, it could be the only solution in a SaaS service model and also probably in a PaaS model. However, both sets of authors agreed in saying that forensics capabilities in these service models are immature and still need further research to be developed.

### 4.2 Static, Elastic and Live Forensics

Recovering deleted data and identifying its owner in order to use them in an event reconstruction, represent a challenge in cloud. This is because once data is deleted the space is removed from mapping and marked as available for being overwritten in a matter of seconds (Ruan, et al., 2011). Birk & Wegener (2011) first mentioned persistent storage as a possible solution for the preservation of these volatile data. However, it was Zawoad and Hassan (2013) who first provided with a guideline for the procedure based on Birk & Wegener (2011) findings. They mention two possible scenarios; in the first one the users would have available a tool which allows them to preserve the data continuously synchronized in any cloud storage. In the second scenario, the user is the person who has committed the crime, a situation in which he probably would not have interest in using this tool for preserving the evidence. In this case, it would be the CSP who needs to manage the mechanism and preserve the data in their infrastructure. Finally, the combination of using this tool in both parts would allow the forensics investigator to compare them, guarantee the veracity of the data obtained and facilitate its acceptation in court. The authors recommend the use of this technique in IaaS and PaaS services model.

Another problem revealed by Ruan, et al., (2011) came when trying to construct the timeline of an event. They state that is due to both the large number of endpoints and the fact that the data resides in different machines, regions or is flowing between the cloud and the endpoint. In Belorkarar & Geethakumari (2011) the authors suggest a methodology based on regenerating events while doing continuous snapshots, as a solution for this issue. It is also argued by them that the use of this method would result in sequenced and stronger evidence. Recently, other researchers have indicated the use of this technique for regenerating events. Zawoad and Hasan, (2013) refers to that technique as the best option to access all possible kinds of data. Its use has also been indicated in Dykstra and Sherman (2013) where they recommend these techniques as an ideal and necessary complement for FROST.

Regarding the SaaS service model, in Zawoad and Hasan, (2013) it is said that for the time being it is only the provider who has access to the system logs. The authors state that the only way to reduce the level of dependency with the CSP is to implement an API in order to make the logs available for an external investigator.

## 4.3 Evidence Segregation

Although different clients of a cloud are isolated from each other they are using different instances running all in the same machine and hence, sharing the system audit logs. Which is the reason why supposes a challenge segregate the information without compromise the confidentiality of other tenants (Ruan, et al., 2011). The implementation of the tools addressed in the previous section, either FROST or the API system logs, could partially solve the problem, as the applications would keep track of the system logs for itself; it would not be necessary to access the node logs, hence, the investigation would not compromise the data of other tenants. Unfortunately, this solution could be only adopted by SaaS and partially by PaaS as it has been seen in the previous section. (Zawoad and Hasan, 2013) (Dykstra and Sherman, 2013).

On the other hand, apart from segregating the system logs, when conducting a cloud-based investigation, in a much earlier step, the instance which is being investigated needs to be isolated in order to prevent evidence from contamination. However, cloud design makes it difficult because more than one instance is allocated in the same node, making it likely to lose information if the CSP shot down the machine and trys to move it to an isolated node. In this way, some isolation techniques, (Failover, Server Farming, Relocation, MITM, Address Relocation, Sandboxing and LHFTB), were presented in Delport, Köhn & Olivier (2011). Although, their conclusion mentions that none of them can individually comply with all the possible scenarios and requirements needed for a cloud-based investigation, it also says that a combination of more than one technique could be a feasible method to isolate an instance.

## 5. Conclusion and Future Work

It is clear that recently cloud computing has experienced a rapid growth, and academic research has focus on its development recent years However, it has also been seen how security still remain the biggest obstacle to its adoption (Sinha and Khreisat, 2014). On the other hand, although research has shown that there is an urgent need to solve several issues and to develop forensics capabilities in cloud-based environments (Taylor, et al., 2011) (Ruan, et al., 2011)

(Zawoad and Hasan, 2013), according to CSA, 2012 cloud computing is still immature and its maturity is expected in the next decade. Hence, there is still much work to do in order to include proactive countermeasures in the cloud architecture to enhance forensics capabilities for cloud computing.

In this paper, it has been reviewed and referred some of the solutions that different researchers have previously proposed for mitigating or solving the technical challenges posed by cloud computing for a forensics investigation in a cloud based environment; however, the problem is that among them, only FROST has been tested in real conditions. In future we will present and evaluate the implementation of a tool called Forensic Evidence Acquisition and Preservation Tool (FEAP), which based on snapshot techniques, aim to perform the data collection process without the cloud provider assistance.

Even though, cloud forensics has had improvements in recent years, there are still challenges that need further research, such as the current need of dependence on CSPs. As Zawoad and Hasan, (2013) point out while introducing either a robust API or management plane in their infrastructure, the CSPs would shift the remote data acquisition responsibility. Other solution it could be the development of either the Trusted Third Party Model proposed or an alternative in which it would be the provider who manages the data collection in a trustable manner.

Finally, since Ruan, et al., 2011 proposed the three dimensional model, there has been addressed different problems which might be included as part of the technical dimension. Several researchers have recently identified timing as an important emerging issue, due to the almost infinite storage capacity offered by Cloud Computing (Trenwith and Venter 2013) (Thethi and Keane, 2014). Thethi and Keane, 2014 add that further research needs to be done to quicken the forensics investigation process in order to comply with deadlines.